\documentstyle[aps,epsf,epsfig]{revtex}
\def\beq{\begin{equation}}
\def\eeq{\end{equation}}
\def\beqa{\begin{eqnarray}}
\def\eeqa{\end{eqnarray}}
\def\l{\left}
\def\r{\right}
\def\bdi{\begin{displaymath}}
\def\edi{\end{displaymath}}

\begin{document}
\tightenlines

\title{Phase diagram of force-induced DNA unzipping in exactly solvable models}

\author{D. Marenduzzo$^{1,*}$, A. Trovato$^2$, and A. Maritan$^1$}
\address{$^1$International School for Advanced Studies (SISSA),\\
and Istituto Nazionale di Fisica della Materia,\\
Via Beirut 2-4, 34014 Trieste, Italy }

\address{$^1$The Abdus Salam International Center for Theoretical Physics,\\
Strada Costiera 11, 34100 Trieste, Italy}

\address{$^2$Niels Bohr Institutet, Blegdamsvej 17, 2100 K{\o}benhavn {\O}, Denmark}

\maketitle

\begin{abstract}

The mechanical separation of the double helical DNA structure induced
by forces pulling apart the two DNA strands (``unzipping'') has been
the subject of recent experiments. Analytical results are obtained within various models of interacting
pairs of directed walks in the $(1,1,\ldots,1)$ direction on the hypercubic lattice, and the
phase diagram in the force-temperature plane is studied for a variety of cases.
The scaling behaviour is determined at both the unzipping and the
melting transition.
We confirm the existence of a cold denaturation transition
recently observed in numerical simulations: for a finite range of
forces the system gets unzipped by {\it decreasing} the
temperature. The existence of this transition is rigorously
established for generic lattice and continuum space models.

\end{abstract}


\section{Introduction}

In recent years, micro and nanomanipulation of single biological
macromolecules has become feasible due to a dramatic improvement of
experimental techniques. By using devices such as optical tweezers \cite{Svo,Ash},
soft microneedles \cite{Kis} and atomic force microscopes \cite{Han}, it is now possible
to access physical and mechanical properties of fundamental biological
objects, namely proteins, nucleic acids, and molecular motors, on the
scale of individual molecules. Special effort has been devoted to the
measurement of force-elongation characteristics of double stranded DNA
molecules (dsDNA), determining its response to external forces and
torques in the absence of enzymes. The {\it mechanical unzipping} of
dsDNA structure by a force pulling the end of one of the two strands,
the end of the other strand being anchored to some physical support, has
been studied by Bockelmann {\it et al.}\cite {Boc1,Boc2}, who have measured the
average force along the opening of the two strands.
Mechanical forces are in fact exerted on the DNA molecule by different
enzymes during the process of DNA replication or transcription
\cite{Yin,Korn}. On the other hand, the double helical
structure of dsDNA may be disrupted 'in vitro' by changing pH, solvent
conditions and/or temperature \cite{War}. This transition is known as {\it
melting denaturation}, and it has been long studied by theoretical
physicists \cite{Zimm,PS,PeyBi}. Instead, only very recently DNA mechanical
denaturation has been the subject of theoretical studies.
Most of these studies have
considered so far a simple extension of the Poland-Sheraga
model \cite{PS}, in which the two DNA strands are homogeneous ideal
polymer chains interacting with each other, introducing a constant
force pulling apart the two strands by acting on one extremity of both
strands. By using a mapping into a quantum mechanical problem, it has 
been shown \cite{Bha,Nel,Seb,Zhou} that the opening of the two strands occurs
only if the pulling force is increased to a critical value.
The unzipping transition turns out to be a first order phase
transition, whereas the melting transition is second order, in the
ideal case, in $d=3$ (recently both simulations and exact results have
shown that the melting transition becomes first order, when considering
mutually self-avoiding walks (SAWs) \cite{GB,Pel}). The heterogeneous case has
also been studied with similar techniques \cite{Nel}. 
Very recently, Monte Carlo simulations of interacting pairs of self
avoiding walks on a cubic lattice and of bead-and-string chains in the
continuum space have determined the whole phase diagram in the
force-temperature plane, revealing the existence of a re-entrant
zipping-unzipping transition by decreasing the temperature for a
finite range of forces \cite{Amos}.

In this paper, we obtain exact analytical results for a class of
simple lattice models of interacting pairs of homogeneous directed
self-avoiding walks. 
We extend the model introduced in ref. \cite{Amos} in $D=1+1$ to
the generic dimension case $D=d+1$, and analyze both the scaling
behavior of the first order unzipping transition and the
multicritical scaling laws at the melting transition
in the force-temperature plane. We also consider different
versions of the model depending on whether the crossing of the two walks is
allowed or not, or simply penalized by an entropy cost. In all cases,
we find that the critical pulling force {\it increases} with
temperature at low $T$, implying thus the existence of a {\it cold}
unzipping transition \cite{Amos}. This seemingly paradoxical property
is due to the competition between the energy gained by increasing the
open portion of DNA and the entropy lost with the full stretching of
the separated strands. Cold unzipping will be exactly proved for both
ideal and self-avoiding chains in the lattice, and for a discrete
chain with constant distance between consecutive beads in the
continuum space.
We also discuss the role of denatured {\it bubbles} forming in
dsDNA opening, as opposed to the end-opening of the strands induced by
the pulling force. By comparing the phase diagram of the different
models considered, we point out that the approximation of neglecting
bubbles and considering only Y-shaped configurations shares many of the 
features of a mean-field type approximation, with an upper critical
dimension $d_c=4$.

Our paper is organized as follows. In Sec.II we introduce the models of directed walks on the lattice in any dimension $D$ and compute their thermodynamical properties, deriving the phase diagram in the force-temperature plane. In Sec.III we analyze the scaling laws at both the thermal melting and mechanical unzipping transition, and we highlight the features that are physically more interesting in our exact results (namely the existence of a re-entrant cold unzipping transition, our main result, and the role of forbidding the mutual crossing of the two strands). In Sec.IVa we discuss the physical meaning of re-entrance in our (lattice) models, generalizing the proof of its occurrence to the more realistic case of SAWs. In Sec.IVb we prove that also a discrete chain in the continuum {\it space} undergoes cold unzipping, and then argue why such an effect had not been observed in previous calculations \cite{Nel,Seb,Zhou}, performed in the limit of a continuum {\it chain}. Eventually, we discuss some related models (with generic SAWs) and give some additional technical details in the Appendices.

\section{Exact solution of the lattice models}

\subsection{Introduction of the models and their behaviour for thermal melting}

We consider a simple class of models in a $D=d+1$ dimensional hypercubic lattice
$\mathbf{Z^D}$. The two strands of a homogeneous
DNA molecule with $N$ base pairs are mimicked by two SAWs, directed along the $(1,\ldots,1)$ direction. The two
chains have one end in common, while at the other end a force $\vec g$
is pulling in the $(1,-1,0,\ldots,0)$ direction. The walks gain a binding
energy $-\epsilon$ ($\epsilon>0$) every time the bases with the same monomer index (or same projection upon
$(1,\ldots,1)$) interact, i.e. {\it wrong base pairing}
is forbidden.

The canonical partition function for two $N$-step
directed self avoiding walks (DSAWs) is: 

\beq\label{1}
{ Z_{N}(\beta,\vec g)=\sum_{\vec x\in \mathbf{Z^D}}p_N(\vec
x,\beta\epsilon)\exp({\beta \vec g\cdot\vec x}) }, 
\eeq 
where $p_N(\vec x,\beta\epsilon)$ is the canonical partition function of two
directed interacting strands whose last base pairs are at relative
distance $\vec x$. The following recursion relation holds for
$p_N(\vec x,\beta\epsilon)$:

\beq\label{2}
p_{N+1}(\vec x,\beta\epsilon)=\sum_{i,j=1}^D p_N(\vec x - \vec e_i + \vec
e_j,\beta\epsilon)(1+(\exp(\beta\epsilon)-1)\delta_{\vec x,\vec 0})\:, 
\eeq 
where $\vec e_i$, $i=1,\ldots,D$, are the canonical euclidean versors
of the $D$-dimensional space and $\delta$ is the Kronecker
delta.

These and similar equations have been intensively studied, at
$\vec g = \vec 0$, within simple models of DNA thermal melting \cite{GB,Amos}, and,
in $D=2$, within the context of random walk adsorption \cite{R1} and
wetting problems \cite{For}. 
Note that the choice of the $(1,\ldots,1)$ direction along which the walks are directed, is crucial in allowing us to write local recursion relations. It is thus no surprise that the model belongs to the same universality class of random walks in $d = D-1$ dimensions \cite{Car}. 

It is well known (see \cite{GB} for a recent review), that,
in the absence of a pulling force, within the model defined by
equation (\ref{2}), 
dsDNA undergoes a phase transition between a low temperature double
stranded phase and a high temperature denaturated state only for $D>3$
($d>2$): in $D=2,3$ dsDNA remains double stranded at all temperatures.

However, we note that in (\ref{2}) the two strands are allowed to
cross each other, without any restriction.
This seems rather unphysical, since every real chain has a
finite "hard core" distance. Thus, one should consider also the case in
which crossing between different strands is forbidden (recent studies
of both homogeneous \cite{Per} and heterogeneous \cite{Hwa} DNA
melting transition have indeed considered the $d=1$ case with
forbidden crossing). Let us focus for the moment on the $D=2$ case, with no 
pulling force, where the effect of forbidding crossing is most
dramatic. Here, the direction $(1/\sqrt{2},1/\sqrt{2})$ can be
identified as "time", and its normal $(1/\sqrt{2},-1/\sqrt{2})$ as
"space" ($x$). The model with crossing (w.c.) implies no restriction
on the relative distance $x$, whereas the one without crossing (w.o.c.) implies that $x$ cannot change sign, {\em e.g.} $x\ge 0$. The
model w.o.c. is equivalent to surface adsorption models previously
considered \cite{R1}. While the thermal melting of the two strands
w.c. takes place at $T_c=\infty$, in the model
w.o.c. $T_c=\epsilon/\log\l(4/3\r)$ as it can be deduced from ref.\cite{R1}.
To elucidate this point further,
we have tackled an intermediate model, where we do
not forbid crossing, but we make it disadvantageous, by assigning a cost $V>0$ each time the strands pass through one another. With calculations similar to those reported below, we find that
melting takes place at the critical temperature (as a function of
$V$): 

\beq\label{3}
T_c(V) = \epsilon \l[\log\l(\frac{4e^{V}}
{3e^{V}+1}\r)\r]^{-1}.
\eeq
As expected, as $V\to \infty$, equation (\ref{3}) yields $T_c\to
\epsilon/\log\l(4/3\r)$, whereas $T_c\sim\epsilon/V$ as $V\to 0$. 

To implement the non-crossing constraint in the model in generic dimension,
we require that, if the DSAWs join, one coming from direction $\vec
e_i$ and the other from direction $\vec e_j$ (with $i \ne j$), when
they divide again, they cannot both proceed along their previous
direction; i.e we forbid that the first walk goes along direction
$\vec e_i$ {\it and} the second one along $\vec e_j$. This excludes
one configuration out of $D(D-1)$ at the splitting point (the
remaining $D$ possibilities would lead to the zipping of the two strands). 
%
As regards the $D=3$ thermal melting, again the model w.c. has
$T_c=\infty$, while, if crossing is forbidden as described above, the
critical temperature is $T_c=\epsilon/\log\l(9/8\r)$. This
result was also found in a similar calculation by Rubin \cite{Rub2}.
For $D>3$ the models with and without crossing both undergo a
denaturation transition at a finite, though different, critical
temperature, so that the effect of forbidding crossing here is less
important (the critical temperatures are the same at the leading order
$1/D$ in the $D \to \infty$ expansion, as expected).

\subsection{Behaviour of the models at non zero force} 

Let us now turn to the models with the pulling force $\vec{g}$ in generic
dimension $D$, with $\vec{g}=(g,-g,0,\ldots,0)$. We can find the
asymptotic value of the canonical partition function (\ref{1}) by
locating \cite{GB,Amos,Lif} the singularity closest to the origin of its
generating function:

\beq\label{GP}
 Z(z,\beta \epsilon,\beta g)=\sum_{N=0}^\infty z^N \sum_{\vec x}
 p_N(\vec x,\beta\epsilon) \exp({\beta \vec g \cdot \vec x}), 
\eeq

which we will also refer to as the grand partition function. We stress
that $p_N(\vec x,\beta\epsilon)$ is a generic partition function: its detailed form
will depend on whether or not we allow crossing. 

To proceed, it is useful to partition the DNA molecule in ds helices
(with the strands attached to each other) and {\it bubbles}, sequences
in which the two chains share just the first and the last base pairs;
we also single out the contribution of the unzipped end of the two
DSAWs, {\it i.e.} the part from the last contact to the end. In this
way, the grand partition sum (\ref{GP}) can be expressed as:

\beq\label{CONV}
Z(z,\beta \epsilon,\beta g)=\frac{1}{1-Dz\exp({\beta \epsilon})}
\frac{\exp({\beta\epsilon})}{1-\frac{\exp({\beta\epsilon})}{1-Dz\exp({\beta
\epsilon})}B(z)} S(z,\beta g),
\eeq
where we have defined:

\beqa\label{def1}
B(z) = \sum_{N=2}^\infty z^{N}b_{N}, \,\
S(z,\beta g)= 1+\sum_{N=1}^\infty z^{N} \sum_{\vec x} \l[c_{N-1}(\vec x)
- c_{N-1}(\vec 0)\delta_{\vec x,\vec 0}\r] \exp({\beta \vec g \cdot \vec
x}). 
\eeqa
In (\ref{def1}), $b_N$ is the number of $2N$-step bubbles,
and $c_N(\vec x)$ the partition function of two $N$-step DSAWs never touching
each other and having their last sites at a mutual distance $\vec x$,
and their initial sites at a relative distance $\vec e_{i_0} - \vec e_{j_0}$
for some $i_0 \ne j_0$. By summing over all possibilities for initial
conditions (see equation (\ref{SC}) 
 below), $b_N=\sum_{i \ne
j;i,j=1}^D c_{N-2}(\vec e_i - \vec e_j)$, so that we need to find an
explicit expression for the $c_N$'s. The equations they obey are: 
\beq\label{RecRel}
c_{N+1}(\vec x)  = \sum_{i,j=1}^D c_N(\vec x-\vec e_i + \vec e_j)- c_N(\vec 0)\sum_{i\ne j;i,j=1}^D \delta_{\vec x, \vec e_i-\vec e_j}.
\eeq
Notice that in eq.(\ref{RecRel}) $\vec 0$ acts as a
sink or absorbing state, {\em i.e.} once the two walks
join, they never leave. In this way, $c_N(\vec x)$ with 
$\vec x\ne\vec 0$ counts
the number of pairs of walks that never touch each other
(and with the last monomers at a relative distance $\vec x$),
while $c_N(\vec 0)$ counts the number of remaining pairs of
walks that at some point come into contact and then remain
stuck together.
This last quantity plays the role of an arbitrary constant and has to be subtracted away from the final counting as in (\ref{def1}) (see \cite{Rub2} for another example in which this prescription was used in a $D=3$ example). 

 In order to simplify the presentation from now on
we restrict our calculations on models w.o.c. 
(detailed calulations for this case and the outline of calculations for the simpler models w.c. are deferred to Appendix A).
The initial conditions for such models are:

\beqa\label{SC}
c_{0}(\vec x) & = &\l(\sum_{i \ne j;i,j=1}^D \delta_{\vec x,\vec e_i -
\vec e_j}\r) \; - \; \delta_{\vec x,\vec e_{i_0} - \vec e_{j_0}}, 
\eeqa
where $\vec e_{i_0}$ and $\vec e_{j_0}$ are the directions forbidden
as described in the previous subsection. 

 By performing a Fourier and a discrete Laplace transform
on eq.(\ref{RecRel}), and by using the relation between
$b_N$ and $c_N$ given above, we can derive an explicit expression
for the bubble generating function $B(z)$ (see eq. (\ref{B(z)})).
The singularities of the
second denominator of the grand partition function (\ref{CONV}) are
necessary to find the melting temperature. We find that 
the equations locating these two singularities are :
\beqa\label{z_1}
 z_1 & = & \frac{1}{D^2},
\\ \label{z_2}
\exp({-\beta\epsilon})-Dz_{2} & = & B \l(z_2\r).
\eeqa
The first singularity $z_1$ is
leading to the usual random walk behavior in the absence of any
interaction.
The second singularity $z_2\l({\beta\epsilon}\r)$ is a
function of the strength of the attractive interaction between the two
strands, determining thus the behaviour in the native zipped phase. At
$g = 0$, the critical temperature for thermal melting is
obtained when the two singularities above coincide ($z_1=z_2$). 
Unlike the models w.c., which have
$T_c=\infty$ in $D\le 3$,  those w.o.c. have a finite critical temperature in
any $D\ge 2$ (see Appendix A for further details).

As regards the force-dependent third factor in (\ref{CONV}), it reads:
\beq\label{CT}
S(z,\beta g) = zS_{sing}(z,\beta g) + 1 - z c(z,\vec 0).
\eeq
 A third new singularity, depending on the external force,
and fundamental in our calculations arises
when computing:
\beq\label{S_sing}
S_{sing}(z,\beta g)\equiv \int_{[-\pi,\pi]^D}\frac{d^D \vec q}{(2\pi)^D}
\stackrel{\sim}{c}(z,\vec q) \sum_{\vec x} \exp[{(\beta\vec g - i\vec
q) \cdot \vec x}], 
\eeq
where $\stackrel{\sim}{c}(z,\vec q)$ and $c(z,\vec 0)$ are given in eqs. (\ref{partialresult},\ref{part2}).
Notice that in eq.(\ref{S_sing}) one can immediately compute the 
integrals on $dq_3\ldots dq_D$, by using the well known identity
$2\pi\delta\l(q\r)=\sum_x\exp{\l(- i q x\r)}$, so that one is left 
with the double integral:
\beq\label{integral}
\int_{-\pi}^{\pi}\frac{dq_2}{(2\pi)}\int_{-\pi}^{\pi}\frac{dq_1}{(2\pi)}
\stackrel{\sim}{c}(z,q_1,q_2,0,\ldots,0)\sum_{x_1,x_2} \exp[{(\beta  g - i
q_1) x_1+(-\beta  g - iq_2)x_2}]. 
\eeq
We now give details for the evaluation of the inner integral only, the outer 
one being equivalent. Defining $h(q_1,q_2)\equiv \l(\stackrel{\sim}{c}(z,q_1,q_2,0,\ldots,0)\r)^{-1}$,
we aim at proving that the complex translation $q_1\to q_1-i\beta g$ can be performed in the integral
$\int_{-\pi}^{\pi}\frac{dq_1}{2\pi}\frac{\exp{\l(-iq_1 x_1\r)}}{h(q_1,q_2)}$.
Notice that $h$ is a periodic function of $q_1$. 
We extend $q_1$ to complex numbers and consider the
contour $\gamma$ as shown in Fig. 1. By the residue
theorem, one can write:
\beqa\label{residue}
\int_{-\pi}^{\pi}\frac{dq_1}{2\pi}\frac{\exp{\l(-iq_1 x_1\r)}}{h(q_1,q_2)} & + &
\int_{0}^{-\beta g}\frac{i dy}{2\pi}\frac{\exp{\l(-i\pi x_1+x_1 y
\r)}}{h(\pi+iy,q_2)}-\\
\nonumber\int_{-\pi}^{\pi}\frac{dq_1}{2\pi}\frac{\exp{\l(-iq_1 x_1-\beta g x_1\r)}}{h(q_1-i\beta g,q_2)} & + &
\int_{-\beta g}^{0}\frac{idy}{2\pi}\frac{\exp{\l(i\pi x_1+x_1 y\r)}}{h(-\pi+iy,q_2)}=\sum_{q_0}{\rm Res}({\cal{H}},q_0),
\eeqa
 where the sum runs over all poles $q_0$ of ${\cal{H}} \equiv\frac{\exp{\l(-iq_1x_1\r)}}{h(q_1,q_2)}$ (at fixed $q_2$)
inside the contour $\gamma$ and ${\rm Res}({\cal{H}},q_0)$ is the residue of ${\cal{H}}$ at $q_0$.
Note that due to the periodicity of $h$ the second and fourth terms
in the previous equation cancel and thus the complex translation can be performed provided that no pole 
of ${\cal{H}}$ can be found inside the contour $\gamma$.
An equivalent conclusion
can be reached for the outer integral on $dq_2$.
The condition of having no poles inside the contours of integration
is satisfied as long as $z<z_3(\beta g)$, where: 
\beq\label{z_3} 
z_3(\beta g)=\frac{1}{\l(D-2\r)^2 + 2 + 2\cosh\l(2\beta g\r) +
4\l(D-2\r)\cosh\l(\beta g\r)}
\eeq
is found by solving $h(-i\beta g,i\beta g) \propto (1-zf(-i\beta\vec g))=0$, with $f(\vec q)$ defined in eq. (\ref{def2}).
In this case, we find consistently that $z_3(\beta g)$ is just the
singularity in the resulting expression $S(z,\beta g) = 1 + z \l[
\stackrel{\sim}{c}\l(z,-i\beta\vec g\r) - c(z,\vec 0)\r]$.

As $z_3$ is always smaller than $z_1$ (for $g\ne 0$), the singularity
closest to the origin has to be determined between $z_2$, controlled
by the attractive energy $\epsilon$, and $z_3$, controlled by the
pulling force $g$. If $z_2 < z_3$, the DNA molecule is zipped,
otherwise it is unzipped. The free energy per monomer $f$  and the average distance between the two
ends (projected onto $\hat{g}\equiv (1,-1,0,\ldots,0)$), $<x_g>$, are defined as:

\beqa\label{def_s}
& & f  \equiv
\lim_{N\to\infty}-T\frac{\log\l[\sum_{\vec x} p_N(\vec x,\beta\epsilon) \exp\l({\beta
\vec g \cdot \vec x}\r)\r]}{N}\\
& & < x_g > \equiv \lim_{N\to\infty}<{\hat{g}}\cdot{\vec x}> = -\frac{\partial
f}{\partial g} N
\eeqa
These quantities read in the thermodynamic limit:

\beqa\label{fe}
& f & = T\log{z_2\l(\beta \epsilon\r)}\:; \,\
\frac{<{x_g}>}{N} = 0 \qquad g < g_c(T,\epsilon) \\
& f & = T\log{z_3\l(\beta g\r)}\:; \,\ \frac{<{x_g}>}{N}
=  z_3\l(\beta g\r) \l[ 4\sinh\l(2\beta g\r) + 4\l(D-2\r)\sinh\l(\beta
g\r)\r] \quad g > g_c(T,\epsilon) 
\eeqa
where the critical force $g_c(T,\epsilon)$ is found by imposing
$z_2\l(\beta \epsilon\r) = z_3\l(\beta g\r)$, as given in eqs.
(\ref{z_2},\ref{z_3}). The above equations show the existence of a
first order transition at $g=g_c(T,\epsilon)$, if
$g_c(T,\epsilon)>0$.  The phase diagram for the models w.c. can be
found exactly in the same way (see Appendix A for some details on this). It is interesting to notice that the singularity $z_3(\beta g)$ does not 
depend on whether or not we allow crossing, whereas $z_2$
in the w.c. case is different from the w.o.c. case. This is enough
of course to make the whole phase diagram different for the two kinds of 
models, as will be discussed in section III.

We stress here that the main interest in using directed walks
is that they are a sub-class of SAWs in the same dimension.
However, qualitatively similar results can be obtained also with
simple random walks (RWs)\cite{rw_note}, but in that case
it would be not physically meaningful to forbid crossing as done above.

Experimentally, however, it should not be hard to set up an 
experiment closely related to the calculation we have performed, by 
stretching the ds molecule in one given direction before applying the
external force.

\section{Scaling laws for thermal melting and unzipping}

Let us focus for concreteness on the result for the model in $D=2$
w.o.c., in order to analyze the scaling laws of the system. 
The phase diagram, as found also in \cite{Amos}, explicitly reads (see
Fig. 2):
\beq\label{pdwoc}
g_c(T,\epsilon)=\frac{T}{2}\cosh ^{-1}{ \l[\frac{1}{2}\frac{1}{\sqrt{1-\exp{(-\beta
\epsilon)}}-(1-\exp{(-\beta \epsilon)})}-1 \r]}.
\eeq 
The model exhibits a first-order ``unzipping'' transition if we move at a fixed
value of $T<\epsilon/\log\l(4/3\r)$, as shown in (\ref{fe}). As is
shown in Fig. 2, $\lim_{T\to 0}g_c(T,\epsilon)=\frac{\epsilon}{2}$ and $g_c(T,\epsilon)$ attains its maximum at $T=T_M\simeq0.9\epsilon$ where $g_c(T_M)\simeq0.68\epsilon>\frac{\epsilon}{2}$.  
The transition is second order at 
$g=0$; as in \cite{GB} we find that close to the
melting point 
\beq\label{new1}
<n>\equiv\frac{\partial}{\partial (\beta \epsilon)}
\log{Z_N}\sim\tau^{-1} f\l(\tau N^{\phi_t}\r),
\eeq
where $<n>$ is the mean number of native contacts,
with $\tau=\l(T-T_c\r)/T_c$, and $\phi_t = 1/2$ for the crossover
exponent at thermal melting in the $D=2$-case. The scaling function $f(x)$ 
behaves as usual 
\beq\label{f(x)}\l\{\begin{array}{ll}
f(x) \to   0 & \qquad {\rm \, as \, } x\to +\infty,\\
f(x) \sim  x & \qquad {\rm \, for \, } |x|\ll 1,\\ 
f(x) \sim  -\l|x\r|^{1/\phi_t} & \qquad {\rm \, for \, }  x\to -\infty,
\end{array}\r. \eeq
 such that $<n>\sim
N^{\phi_t}$ at the transition. 

The exact expressions for $<n>$ and for $<x_g>$ can be found by
inverse Laplace transforms of the 
quantities $Z(z,\beta g,\beta \epsilon)$, $\frac{\partial}{\partial
(\beta\epsilon)}Z(z,\beta g,\beta \epsilon)$,$\frac{\partial}{\partial
(\beta g)}Z(z,\beta g,\beta \epsilon)$. However, if we only need scaling
relations in the thermodynamic limit, we can use the discrete Tauberian theorem
(see \cite{H} and Appendix B), which relates the critical behaviour of a series to the
asymptotic behaviour of its coefficients. By using this method, we
find, in the vicinity of the unzipping mechanical transition:
\beqa\label{new2}
& &<n> \sim
\frac{h_1\l[N^{\phi_1}\l(A_1\gamma+A_2\tau\r)\r]}{A_1\gamma+A_2\tau},
\\ \label{new3} & & <x_g> \sim  - \frac{h_2\l[-N^{\phi_2}\l(A_1\gamma+A_2\tau\r)\r]}{A_1\gamma+A_2\tau},
\eeqa
where $A_1(T_c,g_c),A_2(T_c,g_c)$ are determined in such a way that
$\frac{{\rm d}g_c(T,\epsilon)}{{\rm d}T}=-\frac{A_2}{A_1}\frac{g_c}{T_c}$, $\gamma = \l(g-g_c\r)/g_c$,
$\tau=\l(T-T_c\r)/T_c$, and $\phi_1=\phi_2=1$
for the two crossover exponents, consistently with the fact that the
unzipping transition is first-order (the explicit form of $A_1(T_c,g_c),A_2(T_c,g_c)$ is worked out in Appendix B). Note that $A_1>0$, whereas $A_2$ is negative in the re-entrant part of the transition curve.  The scaling functions $h_{1,2}(x)$ behave in a similar way to $f(x)$, and thus we get that
$<n>\sim N^{\phi_1}$ and $<x_g>\sim N^{\phi_2}$ are both extensive
at the transition point. The physical interpretation is that a
macroscopic portion of the chain is still in the double stranded
state, but the rest of the chain is unzipped; these are just the two
phases coexisting at the first order transition. This result will be used later to justify the use of Y-shaped configurations.

It is instructive to compare this phase diagram with that of the same
$D=2$-case when we allow crossing. The transition line (plotted in
Fig. 2), obtained from $z_2=z_3$ (see Appendix A) is: 
\beq\label{pdwc}
 g_c(T,\epsilon) = T \tanh^{-1}\l[{1-\exp{(-\beta\epsilon)}}\r] =
\frac{\epsilon}{2} + \frac{1}{2\beta}\log\l[{2-\exp{(-\beta\epsilon)}}\r] , 
\eeq
Both the models behave similarly near $T=0$, namely they yield a
transition to the {\it cold denaturated} state described in
\cite{Amos}. When we move at a constant force $g$ such that $g_c(0)<g<Max_{T}\l[g_c(T,\epsilon)\r]$, at low
enough temperatures 
the molecule is unzipped, and it zips by increasing $T$. This feature of the phase
diagrams seems at first sight paradoxical, since one would expect the
critical force to decrease monotonically as the temperature is increased. The physical explanation of this result will be given below. In the model w.c., moreover, $g_c$ always increases,
approaching $\epsilon$ as $T\to\infty$, and the two strands remain
zipped for every temperature when $g<\epsilon$. In the model w.o.c., instead, the two strands unzip again by
further increasing the temperature. As regards scaling, on the other hand, the two models are identical so that we can say that the effect of forbidding crossing is irrelevant in the renormalization group sense, but has a dramatic effect on the form of the critical line.

Let us now discuss the results that we have obtained in higher
dimensions (the critical lines for three and four dimensional models
are shown in Fig. 3). Common to all models is the
cold unzipping transition found in the
$D=2$ case. Moreover, the scaling laws at the unzipping transition
do not depend on dimensionality in the crossover exponents. On the other hand, the thermal melting has
dimensionality-dependent critical exponents (see
\cite{PS,GB}). Consequently, the behaviour of the critical line $g_c(T,\epsilon)$
near the end point $T=T_c$ also depends on $D$. As $T \to
T_c^{-}$ the result that we find in generic dimension is: 
\beqa\label{endpoint}
{\frac{g_c(T,\epsilon)}{T} \sim \left\{\begin{array}{ll}
\epsilon /T -\epsilon /T_c(D) & \textrm{$D=2,4$} \\
\exp\l[-\frac{a}{(\epsilon /T -\epsilon /T_c(D))}\r] & \textrm{$D=3$} \\
\l[\epsilon /T -\epsilon /T_c(D)\r]^{1/2}\l[\log{(\epsilon /T -\epsilon
/T_c(D))}\r]^{-1/2} & \textrm{$D=5$} \\
\l[\epsilon /T -\epsilon /T_c(D)\r]^{1/2} & \textrm{$D>5$}
\end{array}\right.
},
\eeqa
where $a$ is a constant, and $T_c=\infty$ in $D=2,3$ for models w.c.\\
\\

\section{Physical explanation of re-entrance}

\subsection{Re-entrance in lattice models}
To obtain some physical insight into our analytical results,
we compute the free energy of a Y-shaped molecule of
DNA with a force pulling at the extremities (see Fig. 4). 
In this way, we neglect all the configurations with {\it bubbles}.
Such an approximation is valid at low T and yields the exact result
in the limit $T\to0$. Since the configurations of the unzipped part
are weighted by $\exp\l({\beta \vec g \cdot \vec x}\r)$, in this limit
only the completely stretched configuration will contribute to the free
energy for the unzipped part of the Y. In the limit of low
temperature, the free energy is then:
\beq\label{reentrance}
{F(m,N) \sim -(N-m)(\epsilon + T \log{\mu}) -2gm},
\eeq
where $m$ is the number of monomers in the unzipped part, $N$ is the
total number of bases and $\mu$ is the connective constant of a single
walk. Note that eq. (\ref{reentrance}) (with $\vec g=g(1,0,\ldots,0)$) is valid also in the case of
two self-avoiding walks (SAWs), since the connective constant of SAWs
constrained to avoid the fully
stretched unzipped part, is the same as for ordinary SAWs \cite{H,Ham} (for a more detailed discussion on SAWs we refer to Appendix C).
By minimizing with respect to $m$, we find the critical force $g_c(T,\epsilon)$
such that (\ref{reentrance}) is minimum for $m=0$ when $g<g_c(T,\epsilon)$, and
for $m=N$ when $g>g_c(T,\epsilon)$:
\beq\label{reentrance1}
{g_c(T,\epsilon)\stackrel{T\to 0}{\sim}\frac{1}{2}(\epsilon + T \log{\mu})}.
\eeq
This is indeed what we have found in our calculations, and confirms
that re-entrance is a robust feature of lattice models, not depending
on dimensionality or self-avoidance. In other words,
(\ref{reentrance1}) means that, at low $T$, it is more difficult to
open a dsDNA helix as $T$ is increased, because the
energy gain obtained through the unzipping is more than compensated by
the entropy loss, since there is only one possible completely
stretched configuration versus $\mu ^{N-m}$ possibilities for the
double stranded portion of the chain (we will see in the following section
that the entropy loss in the continuum space exhibits a power law correction).
For high enough $T$, on the other hand, also other configurations will
contribute to the open portion, increasing its entropy, and the energy
gain will eventually favour the unzipped state, except that in $D=2$
w.c., where the presence of bubble enhances the entropy of the native
portion at any $T$, as explained below.

We have also calculated the phase
diagram obtained by considering only 
the Y configuration: this amounts to putting the generating function for
bubbles $B(z)=0$ in (\ref{CONV}). As expected, the Y-approximation
gives the exact behaviour (\ref{reentrance1}) in the limit $T\to 0$,
whereas it gets wronger and wronger as $T$ is increased (see Fig. 2 where we plot the two-dimensional case explicitly given by $g_c(T,\epsilon)=\frac{T}{2}\cosh^{-1}\l[\exp{(\beta\epsilon)}-1\r]$ ). For the
critical line near the melting point, this approximation yields $g_c(T,\epsilon) \sim(T_c-T)^{1/2}$ for all $D$,
which from (\ref{endpoint}) is correct only for $D>5$ ($d>4$). This is
because such an approach neglects bubbles, which are expected to be
more and more relevant as $d$ decreases (for $d>d_c=4$ once the two
walks have splitted away, they basically never meet again) and $T$
increases. In this 
view, it is clear that the puzzling behaviour found in the $D=2$ model
with crossing, where $g_c$ increases for all $T$, is due to the growing presence of bubbles: before DNA
can be unzipped, we have to disentangle all the bubbles that are
forming, which is harder and harder as $T \to \infty$. 

\subsection{Re-entrance in the continuum}
We now prove that also discrete chains in the continuum space
undergo a cold denaturation for $T\to 0$. Let us consider two
$N$-monomer chains in
$\mathbf{R^d}$ with no constraint on the directedness,
 with a force $\vec g$ pulling at the extremities and with constant unitary distance between consecutive
monomers. As $\beta\to\infty$, bubbles can be neglected and only
Y-shaped configurations contribute to the partition sum. The partition
function for a Y-configuration then reads: 
\beq\label{PART}
Z_N\l(\beta\epsilon,\beta
g\r)=\sum_{m=0}^{N-1}Z_{N-m}^{z}(\beta\epsilon)Z_{2m}^{u}(\beta g), 
\eeq
where $Z_{N-m}^{z}(\beta\epsilon)=\int_{\mathbf{R^d}}d^{d}\vec x_1
\ldots d^{d}\vec x_{N-m-1} \delta\l(|\vec
x_1|-1\r)\ldots\delta\l(|\vec
x_{N-m-1}|-1\r)\exp{((N-m)\beta\epsilon)}$ is the partition function
of the zipped portion of the strands, and $Z_{2m}^{u}(\beta g)=
\int_{\mathbf{R^d}}d^{d}\vec y_1\ldots d^{d}\vec y_{2m}
\delta\l(|\vec y_1|-1\r)\ldots\delta\l(|\vec
y_{2m}|-1\r)\exp{\l(\beta\vec g\cdot\sum_{i=1}^{2m}\vec y_i\r)}$
is that of the unzipped end.
Chain discreteness is crucial in ensuring the
validity of eq.(\ref{PART}) for $\beta\to\infty$, since when one bubble
has formed, the length of the stretched portion of the chain can be at
most $\sum_{i=1}^{2m-1}\vec y_i=2(m-1)\frac{\vec g}{|\vec g|}$.
The integrals in (\ref{PART})
can be performed in any dimensions, and the evaluation of
$Z_{2m}^{u}(\beta g)$ involves the modified Bessel function of the
first kind of order $\frac{d-2}{2}$. In particular, $Z_{2m}^{u}(\beta
g)\stackrel{\beta\to\infty}{\sim}\l(\frac{2\pi}{\beta
g}\r)^{m(d-1)}\exp{(2m\beta g)}$, and there is a power-law correction with respect to the lattice result. We find that $Z_{N}\l(\beta\epsilon,\beta
g\r)\stackrel{\beta\to\infty}{\sim}\sum_{m=0}^{N-1}\Omega_d^{N-m-1}\exp{((N-m)\epsilon)}Z_{2m}^{u}(\beta
g)$, where $\Omega_d$ is the surface area of the unit sphere in $d$ dimensions. From this, the critical force is easily found:
\beq\label{CONTINUUM}
g_c(T,\epsilon)\stackrel{T\to
0}{\sim}\frac{\epsilon}{2}-\frac{d-1}{2}T\log{\frac{T}{\epsilon}}-\frac{T}{2}\log{\l[2^{2d-3}\pi^{\frac{d}{2}-1}\Gamma
\l(\frac{d}{2}\r)\r]}, 
\eeq
where $\Gamma$ is the Euler Gamma function. The critical
line $g_c(T,\epsilon)$ increases at low $T$ and re-entrance is
present also in the continuum, even enhanced with respect
to the lattice case. The leading term $T\log T$ in equation (\ref{CONTINUUM}) (due to the power-law correction in $Z_{2m}^{u}(\beta
g))$
 is indeed not present in the $d=1$ case, when one correctly
recovers equation (\ref{reentrance1}) for a lattice random walk with $\mu=2$.

We finally wish to discuss the relation of the present treatment with
previous work (\cite{Bha,Nel,Seb,Zhou}) done on the unzipping of homo-DNA
in the limit of a continuum chain.
The ds molecule with the pulling force had been described by means of an effective hamiltonian, which, apart from an irrelevant center of mass term, reads:

\beq\label{eff_ham_}
H=\int_0^N dn\l(\frac{Td}{b^2}\l(\frac{d\vec r}{dn}\r)^2+
V\l(\vec r(n)\r)-\vec g\cdot \frac{d\vec r}{dn}\r),
\eeq

where $\vec r$ is the relative separation between the strands, $b$ is the effective Kuhn length of single-stranded DNA and $V$ is a realistic short range attractive potential, namely a delta function or a potential well (the two cases should be equivalent according to the standard quantum theory as long as $\int d\vec r V(\vec r)$ is the same). The system described by (\ref{eff_ham_}) is equivalently represented by a quantum system with a non-hermitian hamiltonian. One finds \cite{Nel,Seb} that there is a first order unzipping transition when the force reaches the critical value:

\beq\label{critical_force}
g_c(T,\epsilon)=\sqrt{-\frac{4\epsilon_0(T)Td}{b^2}},
\eeq

where $\epsilon_0(T)$ is the ground state energy of the quantum hamiltonian obtained from (\ref{eff_ham_}) when $\vec g=\vec 0$. In the quantum mechanical system, for $g<g_c(T,\epsilon)$ the ground state is a bound state, while for $g>g_c(T,\epsilon)$ the spectrum of (\ref{eff_ham_}) is continuous.
 
Let us focus on the $d=1$ case, corresponding to our two-dimensional
models without crossing constraints. It is well known that at $g=0$
both a symmetric square well and a delta function always have at least
one bound state, meaning that DNA remains ds at all $T$. In Fig. 5 we show the critical force for the two potentials (obtained simply from (\ref{critical_force})). It can be seen from the figure that at high $T$ both the potentials saturate towards the same limit, as in our calculation for the model w.c. This is expected since the delta potential case can be recovered from our equation (\ref{2})  in the limit of both continuum space and time (see Fig. 4). At low $T$, however, $g_c(T,\epsilon)\sim \l(\frac{T}{\Delta}\r)^{1/2}$ for a square well of width $\Delta$, and is constant for the delta potential, so that re-entrance is present only in the former case, but with a behaviour different from that found in the lattice. This already signals that the low temperature behaviour of the solution obtained through the quantum mapping is rather unphysical.
This is due to the fact that in the limit of continuum chain the chain
constraint is modelled in a `soft' way, by using a harmonic potential between
consecutive beads along the chain. This interaction is usually assumed
to be entropic (as in eq. (\ref{eff_ham_})) and thus effectively vanishes
at $T=0$. In that limit eq. (\ref{eff_ham_}) is describing a set of $N$ `free'
 particles moving in an effective potential determined by $V$ and $g$.
Consequently, the quantum mapping is expected to describe the system well except in the low temperature limit. 
Indeed, the re-entrance found for the square well potential is due to
the unboundedness of the potential felt by the ``free particles'' as
soon as $g\ne 0$. As $T\to 0$, fluctuations are not important, and the
particles stay close to the minimum of such potential, which is $-\infty$ for a force whatever small but finite. This is enough to unzip the strands \cite{note_1}.

As Zhou has observed in \cite{Zhou}, one can improve this model by placing a hard core either inside the potential well or at a finite distance from the delta (this is analogous in spirit to introducing the crossing constraint in our lattice models). Once again, one finds that for T near $T_c$, which is now finite, both potentials predict $g_c(T,\epsilon)\stackrel{T\to T_c}{\sim}(T_c-T)$ as in the lattice model (see eq.(\ref{endpoint})), whereas for low $T$ the delta shows no re-entrance and the potential well displays a $T^{1/2}$ behaviour. We note once again that the low temperature behavior of such `quantum' models is an artifact, when considered as polymer chain models, due to the `soft' enforcing of the chain constraint. In this respect, discrete chain models in the continuum space are more realistic, either with constant bond length or with harmonic springs between consecutive beads. In the first case we have indeed proved (see eq. (\ref{CONTINUUM})) the existence of cold mechanical denaturation. In the second case it has been shown to occur by means of numerical simulations \cite{Amos}.

\section{Conclusions}

To conclude, we have studied simple models for DNA mechanical
unzipping induced by a pulling force. By using analytical techniques,
the relevance of forbidding the crossing of the two strands and the
role played by bubble formation in the denaturation process has been
discussed throughout the whole force-temperature plane. The scaling
properties of the system along the transition line have also been
determined. The existence of a {\it cold mechanical denaturation} for
a finite range of forces at low enough temperatures has been exactly
proved in a general case for both lattice and continuum space
models. The fundamental role of chain discreteness has been emphasized
in comparison with related models studied by means of quantum mapping
techniques. Even if we neglected many important effects, such as base pairs
heterogeneity, intrinsic helicity, wrong base pairing, and different
stacking energy for the natured and denatured state, it would be
interesting to experimentally test this prediction stemming from such a
simple model.

Furthermore, on the basis of preliminary results (analytical and numerical) we anticipate that even in the presence of {\em quenched} randomness in the contact potential our models display a re-entrant transition, so in this sense it is really a ``universal'' feature.

{\em Acknowledgements:} We would like to thank F. Seno and S.M. Bhattacharjee for illuminating
discussions. DM also acknowledges INFM funding.
This work was supported by MURST grant.

\appendix

\section{Computation of $B(z)$ in models of DSAWs without crossing and 
of the phase diagram for models with crossing}

 We here first work out the details to obtain an explicit expression
for $B(z)$, defined in eq.(\ref{def1}). Starting from equations (\ref{RecRel}), 
we can perform a Fourier and a discrete Laplace transform to obtain:

\beqa\label{LFT}
\stackrel{\sim}{c}(z,\vec q)\equiv \sum_{N=0}^\infty z^N \sum_{\vec x}
\exp({i\vec q \cdot \vec x}) c_N(\vec x)= \frac{1}{1-zf(\vec
q)}(\stackrel{\sim}{c_0}(\vec q)-zf_0(\vec q)c(z,\vec 0)), 
\eeqa
where we have made the following definitions:
\beqa\label{def2}
& & f_0(\vec q) \equiv 2 \sum_{i<j;i,j=1}^D \cos\l(q_i-q_j\r); \,\
f(\vec q) \equiv D + f_0(\vec q) \\
& & c(z,\vec x) \equiv \sum_{N=0}^\infty z^N c_N(\vec x);\,\
\stackrel{\sim}{c_0}(\vec q) \equiv \sum_{\vec x} \exp({i\vec q \cdot
\vec x}) c_0(\vec x) 
\eeqa
Using $c(z,\vec 0)=\int_{[-\pi,\pi]^D} \frac{d^D \vec q}{(2
\pi)^D}\stackrel{\sim}{c}(z,\vec q)$, equation (\ref{LFT}) is easily
solved since, by permuting variables inside the resulting integrals,
it is possible to see that they do not depend on the particular step
which we forbid. We obtain: 
\beqa\label{partialresult}
\stackrel{\sim}{c}(z,\vec q) & = & \frac{1}{1-zf(\vec
q)}\frac{\stackrel{\sim}{c_0}(\vec q)-zW_1(z)[A(D)f_0(\vec
q)-\stackrel{\sim}{c_0}(\vec q)]}{1+zW_1(z)},\\ \label{part2}
c(z,\vec 0) & = & A(D)\frac{W_1(z)}{1+zW_1(z)},
\eeqa 
where $A(D)=1-\frac{1}{D(D-1)}$ is the reduction factor 
due to the crossing constraint.
As a result, using $B(z)=z^2\sum_{i \ne j;i,j=1}^D c(z,\vec e_i - \vec
e_j)$, we obtain
\beq\label{B(z)}
B \l(z\r) = A(D)\frac{z^2W_2(z)}{1+z_2 W_1(z)},
\eeq
where
\beq\label{wint}
W_1(z)=\int_{[-\pi,\pi]^D} \frac{d^D \vec q}{(2
\pi)^D}\frac{f_0(\vec q)}{1 - zf(\vec q)} \quad , \quad
W_2(z)=\int_{[-\pi,\pi]^D} \frac{d^D \vec q}{(2
\pi)^D}\frac{f_0^2(\vec q)}{1 - zf(\vec q)}.
\eeq
The singularity $z_1=1/D^2$, leading to the usual random walk behavior,
comes in when evaluating the above integrals; their denominator
becomes negative as $\vec q \to 0$ for $z>z_1$.

We give here also a brief outline of the calculation necessary to find out
the phase diagram for the models of Section II of DSAWs
when crossing is allowed. These models are 
simpler to solve than the corresponding models w.o.c., since
it is not necessary to partition the molecule of DNA as done
in the text in eq.(\ref{CONV}), which allows us to write a more explicit 
formula for the grand partition function.
Indeed we can simply start from the recursion 
relations (\ref{2}) and, after Fourier and Laplace transform, find the
expression:
\beq\label{wc1}
\stackrel{\sim}{p}(z,\vec q,\beta\epsilon) = \frac{1}{1-z
f(\vec q)}\frac{1}{1-\l(1-\exp{\l(-\beta\epsilon\r)}\r)W_0(z)},
\eeq
for the quantity $\stackrel{\sim}{p}(z,\vec q,\beta\epsilon)$, defined as:
\beq\label{definition}
\stackrel{\sim}{p}(z,\vec q,\beta\epsilon)=\sum_{N=0}^{\infty}z^N \sum_{\vec x} 
\exp{\l(i\vec q\cdot\vec x\r)}p_N(\vec x,\beta\epsilon)
\eeq
with $p_N(\vec x,\beta\epsilon)$ defined as in eq.(\ref{1}) and
\beq\label{wint0}
W_0=\int_{[-\pi,\pi]^D}
\frac{d^D\vec q}{(2\pi)^D}\frac{1}{1-zf(\vec q)}.
\eeq
The final
form of the grand partition function, for DSAWs w.c. in the
presence of a pulling force, is:
\beq\label{wc2}
Z(z,\beta\epsilon,\beta g)= \frac{1}{1-z/z_3(\beta g)}\frac{1}{1-\l(1-\exp{\l(-\beta\epsilon\r)}\r)W_0(z)},
\eeq
with $z_3(\beta g)$ defined as in eq.(\ref{z_3}).
Alternatively, one can proceed exactly as in Section II,
but with initial conditions 
\beqa\label{SCC}
c_{0}(\vec x) & = &\sum_{i \ne j;i,j=1}^D \delta_{\vec x,\vec e_i - \vec e_j}.
\eeqa
The only resulting difference would be to have
$A(D)=1$ in eqs.(\ref{partialresult}) and (\ref{B(z)});
after a bit of algebra one can convince himself that the partition function
found by recollecting the different factors in eq.(\ref{CONV}) is the same as
that found in eq.(\ref{wc2}). In doing this, one 
needs to compute the relations between the integrals
$W_0(z)$, $W_1(z)$, $W_2(z)$, which may be done as, e.g., in \cite{Rub2}.
For example, the bubble generating function for both kind of models
could also be simply expressed as
\beq\label{B0}
B(z)=A(D)\left(1-Dz-1/W_0(z)\right).
\eeq
Changing the dimension dependent factor $A(D)$ from $A(D)=1$,
when crossing is allowed, to $A(D)=1-\frac{1}{D(D-1)}$,
when crossing is forbidden, is 
enough for the melting transition temperature to become finite for $D\le3$.

\section{Tauberian theorem and its applications}

The discrete Tauberian theorem (see \cite{H}) states that the relations

\beq\label{rel_1}
c\l(z\r)\equiv\sum_{N=0}^{\infty}c_Nz^N\;\stackrel{z\to z^{*-}}{\sim}\l(\frac{1}{1-z/z^*}\r)^{\rho}L\l(\frac{1}{1-z/z^*}\r),
\eeq
and
\beq\label{rel_2}
c_N\stackrel{N\to \infty}{\sim}\frac{N^{\rho-1}}{\l(z^*\r)^N\Gamma(\rho)}L(N),
\eeq
are equivalent provided that $\rho >0$, $\{c_N\}$ is a positive monotonic sequence and $L$ is ``slowly varying'' in the sense specified in \cite{H}.

We apply this theorem here first to find the scaling laws of the unzipping transition to give an example of how the relations reported in section III can be recovered. When approaching the critical line at $g=g_c$ from the unzipped phase, the smallest singularity is $z_3\l(\beta g\r)$, as defined in eq. (\ref{z_3}). The leading behavior as $z\to z_3^{-}$ is:

\beqa\label{unzip_scal_}
Z(z,\beta \epsilon,\beta g) & \stackrel{z\to z_3^{-}}{\sim}
& \frac{1}{\exp({-\beta \epsilon})-Dz_3-B(z_3)}\frac{1}{1-z/z_3},\\ 
 \frac{\partial Z(z,\beta\epsilon,\beta g)}{\partial(\beta g)}&\stackrel{z\to z_3^{-}}{\sim}&
-\frac{1}{\exp({-\beta\epsilon})-Dz_3-B(z_3)}\frac{1}{\l(1-z/z_3\r)^2}\frac{\partial z_3/\partial(\beta g)}{z_3(\beta g)},\\
 \frac{\partial Z(z,\beta\epsilon,\beta g)}{\partial(\beta\epsilon)}&\stackrel{z\to z_3^{-}}{\sim}&
\frac{\exp({-\beta\epsilon})}{\l(\exp({-\beta\epsilon})-Dz_3-B(z_3)\r)^2}\frac{1}{1-z/z_3},
\eeqa
where we have neglected inessential factors when $g\sim g_c$. 
Consequently, the application of the Tauberian theorem yields:

\beqa\label{tauberian}
Z_N(\beta \epsilon,\beta g) & \stackrel{N\to\infty}{\sim}
& \frac{1}{\exp({-\beta \epsilon})-Dz_3-B(z_3)}z_3^{-N},\\ 
 \frac{\partial Z_N(\beta\epsilon,\beta g)}{\partial(\beta g)}&\stackrel{N\to \infty}{\sim}&
-\frac{N}{\exp({-\beta \epsilon})-Dz_3-B(z_3)}z_3^{-N}\frac{\partial z_3/\partial(\beta g)}{z_3(\beta g)},\\
 \frac{\partial Z_N(\beta\epsilon,\beta g)}{\partial(\beta\epsilon)}&\stackrel{N\to\infty}{\sim}&
\frac{\exp({-\beta\epsilon})}{\l(\exp({-\beta\epsilon})-Dz_3-B(z_3)\r)^2}z_3^{-N}.
\eeqa
The canonical partition function $Z_N$ can be expressed in terms of previously defined quantities as $Z_N\l(\beta \epsilon,\beta g\r)=\sum_{\vec x}
 p_N(\vec x,\beta\epsilon) \exp({\beta \vec g \cdot \vec x})$ (see eqs.(\ref{1}) and (\ref{GP})).

Note that at the critical line $z_2\l(\beta\epsilon\r)=z_3\l(\beta g_c\r)$, implying that the $\epsilon$-dependent denominator in the above equations vanishes, as the critical line is approached (that is $\gamma \equiv \l(g-g_c\r)/g_c\ll 1$ and $\tau\equiv\l(T-T_c\r)/T_c\ll 1$). In that limit the order parameters
obey
\beqa\label{n}
& & <n> \equiv \frac{\partial Z_N/\partial(\beta\epsilon)}{Z_N}
\stackrel{N\to\infty}{\sim} \frac{\exp({-\beta\epsilon})}{A_1\gamma+A_2\tau},\\
& & <x_g> \equiv -\frac{\partial Z_N/\partial(\beta g)}{Z_N}
\stackrel{N\to\infty}{\sim} \frac{\partial z_3/\partial(\beta g)}{z_3(\beta g_c)}N.
\eeqa

The coefficients $A_1(T_c,g_c),A_2(T_c,g_c)$ determine the normal direction in the  ($\log(T),\log(g)$) plane, and are reported below:

\beqa\label{A,B}
A_1(T_c,g_c) = & & \beta_c g_c \l(D+\frac{\partial B\l(z_3(\beta_c g_c)\r)}{\partial z_3}\r) z_3^2\l(\beta_c g_c\r) \l[ 4\sinh\l(2\beta_c g_c\r) + 4\l(D-2\r)\sinh\l(\beta_c g_c\r)\r],\\
A_2(T_c,g_c) = & & -\beta_c g_c \l(D+\frac{\partial B\l(z_3(\beta_c g_c)\r)}{\partial z_3}\r) z_3^2\l(\beta_c g_c\r) \l[ 4\sinh\l(2\beta_c g_c\r) + 4\l(D-2\r)\sinh\l(\beta_c g_c\r)\r]+\\ \nonumber
& & \beta_c \epsilon \exp\l(-\beta_c\epsilon\r).
\eeqa
As expected, it may be easily checked that, whatever  direction we
choose to approach the critical curve, the physical quantities as
(\ref{n}) only depend on the projection of this direction on the
normal to the critical line. To show this, one only needs first to
observe that criticality is achieved through the vanishing of the denominator in (\ref{n}), and then to apply the implicit function theorem to this denominator.

Recently, some authors (\cite{Bha,Nel}) have suggested, as an alternative order parameter for the unzipping transition, the number $m$ of monomers from the last contact to the end. By using the same tools as before, we can find some quantities of interest such as the probability distribution of having $m$ monomers ``liberated'' (in the equations below we suppose 1$<<m<<$N). In general:

\beq\label{pot_1}
P(m)=\frac{\sum_W\delta_{n_{u.z.},m}p_W}{\sum_W p_W}
\eeq
where W are the pairs of directed walks, $p_W$ is their Boltzmann weight and $n_{u.z.}$ is the number of ``liberated'' monomers of the configuration. The possible enforcing of the crossing constraint does not affect the following equations, simply shifting the critical melting temperature $T_c$.

At zero force and $T<T_c$, in the native state: 
\beq\label{pot_2}
P(m)\propto\exp{\l[-m\log{(z_1/z_2)}\r]}m^{\alpha(d)},
\eeq
where $m^{\alpha(d)}=m^{-1/2}$ in $d=$1, $\l(\log{(m)}\r)^{-1}$ in $d=$2 and constant for $d>$2. We recall that $z_1 = 1/D^2$, with $D = d+1$.
At criticality:
\beq\label{pot_3}
P(m)\propto(N-m)^{\phi_t-1}m^{\alpha(d)},
\eeq
where in the first factor there are logarithmic corrections in $d=2,4$ to the
power-law behavior controlled by the crossover thermal exponent $\phi_t$ \cite{GB}. Above $T_c$ we have:
\beq\label{pot_3a]}
P(m)\propto(N-m)^{\beta(d)}m^{\alpha(d)}.
\eeq
with $\beta(d)=-|1-d/2|-1$ if $d\ne$2, in $d=$2 $m^{\beta(d)}=1/(N\log^2{(N)})$.
At force different from zero, in the zipped phase, we get:
\beq\label{pot_4}
P(m)\propto\exp{\l(-m\log{(z_3/z_2)}\r)},
\eeq
while in the unzipped phase:
\beq\label{pot_5}
P(m)\propto\exp{\l(-(N-m)\log{(z_2/z_3)}\r)},
\eeq
if $T<T_c$, and
\beq\label{pot_6}
P(m)\propto\exp{\l(-(N-m)\log{(z_1/z_3)}\r)},
\eeq
if $T>T_c$. Just at criticality, the probability distribution is flat.

For the Y-shaped configuration, we get, at force different from zero:
\beqa\label{pot_Y_fne0}
P(m) & \propto & \exp{\l(-(N-m)\log{\l(z_y/z_3\r)}\r)} \quad {\rm u.z. \,\ phase,} \\ \label{fne0_1}
P(m) & \propto & \exp{\l(-m\log{\l(z_3/z_y\r)}\r)} \qquad \quad \quad {\rm z. \,\ phase,} \\ \label{fno0_3}
P(m) & \propto & {\rm const.} \qquad \qquad \qquad \qquad \quad \quad \;{\rm at \,\ criticality,}
\eeqa
where $z_y = \exp\l(-\beta\epsilon\r)/\mu$ is the singularity controlling
the free energy per monomer in the Y-zipped phase, and $\mu = D$ the connective
constant of a single directed walk. At zero force we get:
\beqa\label{pot_Y_fe0}
P(m) & \propto & \exp{\l(-(N-m)\log{\l(z_y/z_1\r)}\r)} m^{\alpha(d)} \quad {\rm denatured \,\ phase,} \\
P(m) & \propto & \exp{\l(-m\log{\l(z_1/z_y\r)}\r)} m^{\alpha(d)} \qquad \quad \quad {\rm native \,\ phase,} \\
P(m) & \propto & m^{\alpha(d)} \qquad \qquad \qquad \qquad \qquad \qquad \;\;\, {\rm at \,\ criticality.}
\eeqa

Let us give the details in a simple case, to get {\em e.g.}  eqns.(\ref{pot_Y_fne0},\ref{fne0_1},\ref{fno0_3}). One has:
\beq\label{newdet}
P(m)=\frac{Z_{N-m}^{PS} Z_{m}^{free}}{Z_N}
\eeq
where $Z_{N-m}^{PS}$ is the partition function of two $\l(N-m\r)$-monomer chains with the end points in common (this is the original Poland-Sheraga model), $Z_m^{free}$ is the partition of two $m$-monomer chains with no base pair in contact, and $Z_N$  is given by eq.(\ref{1}). We have, for the Y at non zero force, $Z_{N-m}^{PS}=z_y^{-(N-m)}$, $Z_{m}^{free}\sim z_3^{-m}$, and

\beqa\label{details}
Z_N & \sim & z_y^{-N} \qquad \qquad \qquad \qquad \qquad\qquad \;\;\, {\rm native \,\ phase,} \\
Z_N & \sim & z_3^{-N}  \qquad \qquad \qquad \qquad \qquad \qquad \;\;\, {\rm above \,\ criticality.}
\eeqa

\section{Y configurations with SAWs}

In this Appendix we extend the analysis of section IV to a Y configuration where the two chains are generic SAWs and not directed walks. We take here $\vec g=g(1,0,\ldots,0)$.

Let us call $c_N(\vec 0,\vec x)$ the number of $N$-step SAWs ending in $\vec x$ and starting in $\vec 0$, and $c_{N,\vec g}=\sum_{\vec x}c_N(\vec 0,\vec x)\exp{\l(\beta\vec g\cdot\vec x\r)}$ (so that $c_N\equiv c_{N,\vec g=\vec 0}$). As a first step, we need to prove the existence of a connective constant $\mu_{\vec g}$ for a single SAW in the presence of a pulling force {\em i.e.} $c_{N,\vec g}\sim \mu_{\vec g}^N$ for large $N$. The existence of a connective constant then follows \cite{H} from the inequality:

\beqa\label{connective}
\sum_{\vec x}c_N(\vec 0,\vec x)\exp{\l(\beta \vec g\cdot\vec x\r)}\le
\sum_{\vec x,\vec x_1}c_{N_1}(\vec 0,\vec x_1)c_{N-N_1}(\vec x_1,\vec x)
\exp{\l(\beta \vec g\cdot\vec x_1\r)}\exp{\l(\beta \vec g\cdot(\vec x-\vec x_1)\r)},
\eeqa
which implies the subadditivity of $c_N$ \cite{H}.
From the connective constants of RWs and DSAWs, we can estabilish the following bounds (recall that here $\vec g=g(1,0,\ldots,0)$)
for the connective constant of a SAW in $d$-dimensions:

\beq\label{bounds}
d-1+\exp{\l(\beta|\vec g|\r)}\le \mu_{\vec g} \le 2(d-1)+2\cosh{\l(\beta|\vec g|\r)}.
\eeq

Let us now introduce the canonical partition function. It reads:

\beq\label{partitionSAW}
Z_N=\sum_{m=0}^{N-1}\exp{\l(\beta (N-m)\epsilon\r)}c_{N-m-1}
\sum_{\vec x_1,\vec x_2} c_{2m}^{\prime}(\vec x_1,\vec x_2)\exp{\l(\beta\vec g\cdot(\vec x_2-\vec x_1)\r)},
\eeq
where $c_{2m}^{\prime}(\vec x_1,\vec x_2)$ counts the SAWs which do not cross the first $N-m$ attached monomers of the Y and $\vec x_1,\vec x_2$ are the end-points of the strands. 

An upper bound is found if we allow the opened part of the Y to cross the first $N-m$ zipped monomers:

\beq\label{part_1_SAW}
Z_N^{upper}=\sum_{m=0}^{N-1}\exp{\l(\beta (N-m)\epsilon\r)}c_{N-m-1}
\sum_{\vec x_1,\vec x_2} c_{2m}(\vec x_1,\vec x_2)\exp{\l(\beta\vec g\cdot(\vec x_2-\vec x_1)\r)}.
\eeq
The grand canonical partition function defined from (\ref{part_1_SAW}) displays singularities in $z\equiv z_1=\frac{1}{\mu_{\vec g}^2}$ and in $z\equiv z_2=\frac{1}{\mu\exp{\l(\beta\epsilon\r)}}$, where $\mu$ is the connective constant of a standard SAW in $d$-dimensional space, {\em i.e.} $\mu=\mu_{\vec g=\vec 0}$. 

To find a lower bound for $Z_N$, we restrict the Y configurations to those with the joined part of the Y constrained to lay in the half-space $\{\vec x|\vec x\cdot\vec e_{1} >0, \vec e_1\cdot\frac{\vec g}{|\vec g|}=0\}$ (the origin is in the bifurcation point), and the unzipped part forced to stay in the other half-space.

\beq\label{part_2_SAW}
Z_N^{lower}=\sum_{m=0}^{N-1}\exp{\l(\beta (N-m)\epsilon\r)}c_{N-m-1}^{left}
\sum_{\vec x_1,\vec x_2} c_{2m}^{right}(\vec x_1,\vec x_2)\exp{\l(\beta\vec g\cdot(\vec x_2-\vec x_1)\r)},
\eeq

where we have indicated with $c_n^{left(right)}$ the number of $n$-step SAWs constrained to stay in the left(right) half space.
As the connective constants can be rigorously proved to be the same even for walks constrained in such a way (see \cite{H,Ham} for details, it is important here that walks generated from a reflection through the hyperplane $\{\vec x|\vec x\cdot\vec e_{1} >0, \vec e_1\cdot\frac{\vec g}{|\vec g|}=0\}$ leave the scalar product $\vec g\cdot\vec x$ unchanged), the singularities of the grand partition function obtained from (\ref{part_2_SAW}) are the ${\it same}$ as those of $\sum_{N=0}^{\infty}Z_N^{upper}z^N$. Thus the grand partition function singularities associated to $Z_N$ in eq.(\ref{partitionSAW}) are $z=z_1,z_2$ given above.

In summary, we have proved that a mechanical unzipping transition takes place when $\mu_{\vec g}^2=\mu\exp{\l(\beta\epsilon\r)}$, similarly to the directed walk case. We suggest that a standard computation of $\mu_{\vec g}$ (for instance through exact enumeration) could give an accurate phase diagram for the ``Y approximation'' with SAWs.

The bounds in eq.(\ref{bounds}) give immediately two bounds within which the transition line $g_c(T,\epsilon)$ lies. In the $T\to 0$ limit eq.(\ref{reentrance1}) follows since both bounds give the same asymptotic transition line. This should be the same also for generic SAWs, since in the $T\to0$ limit it is sufficient to consider only Y-shaped configurations. This demonstrates the existence of cold unzipping at sufficiently low temperature for the generic self-avoiding case.
\newpage
\begin{figure}

\epsfbox{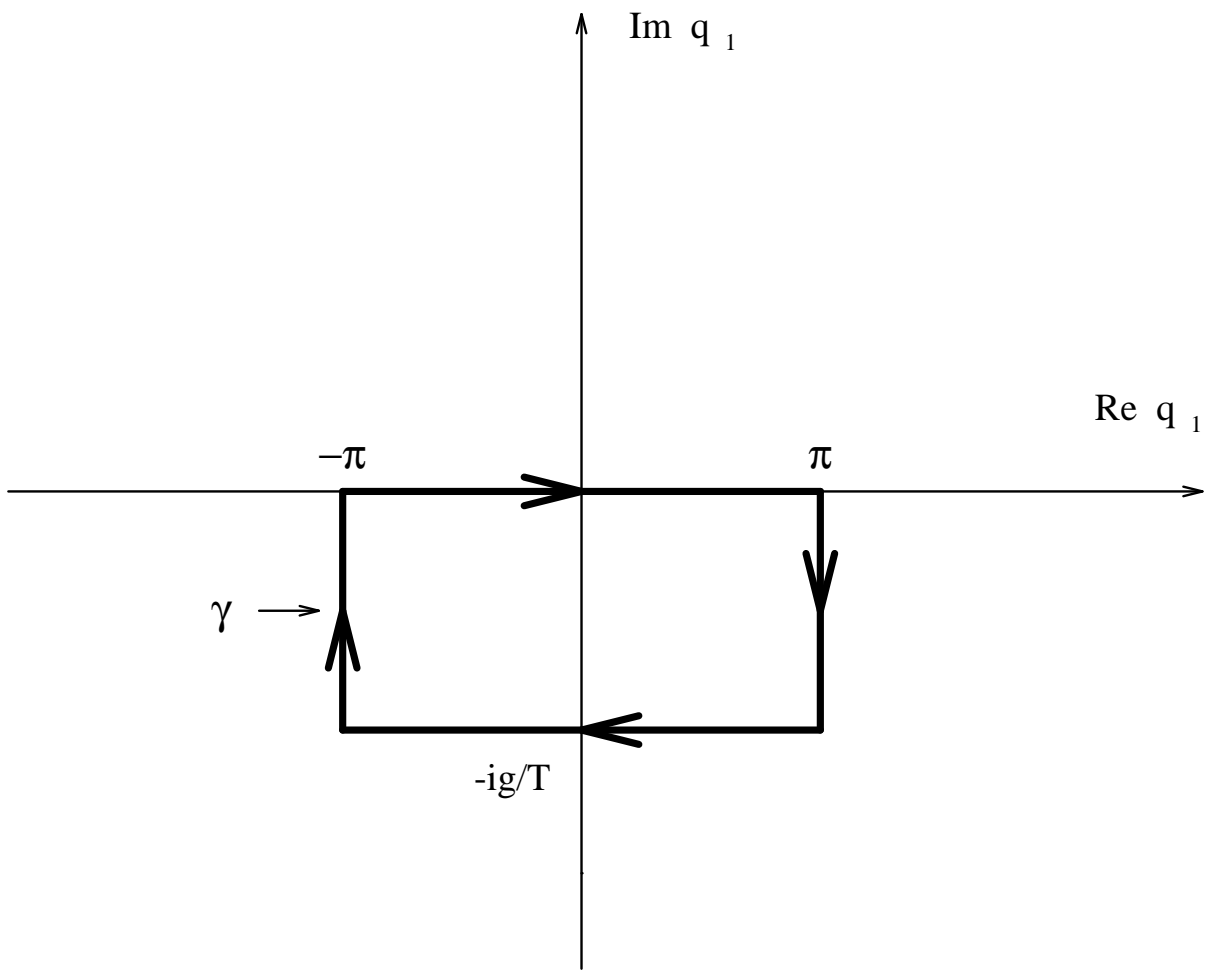}
\\
\caption{Contour of integration $\gamma$ in the complex plane used for the evaluation of the integrals in (\ref{residue}).}
\end{figure} 
\newpage
\begin{figure}\epsfbox{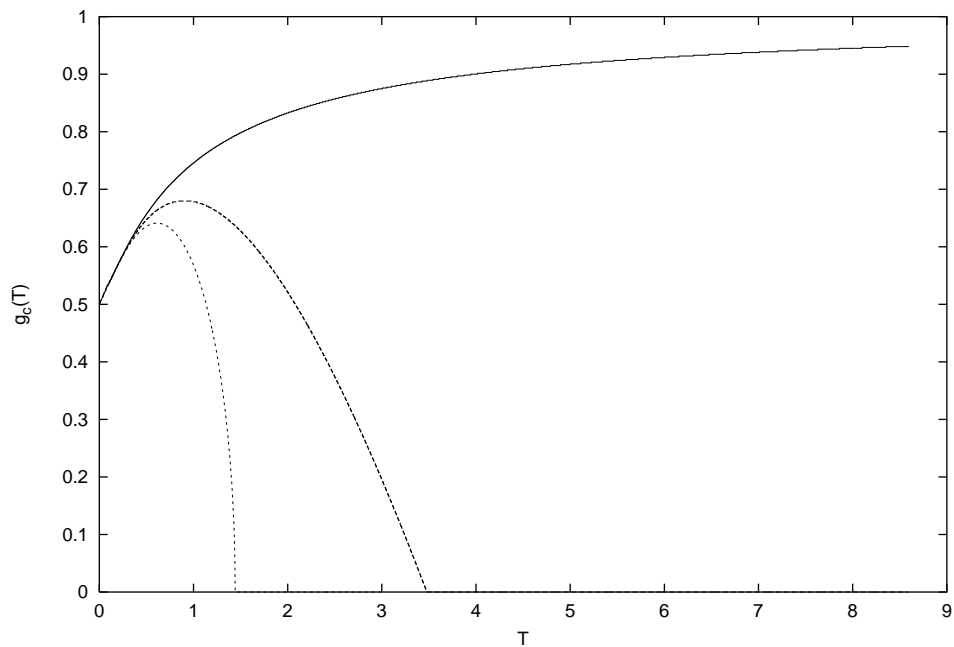}
\\
\caption{Phase diagram for two-dimensional models of DSAWs ($\epsilon=1$ in the figure). The solid curve refers to the model with crossing, the long-dashed one to that without crossing and the dashed one to the model which considers only Y-shaped configurations.}
\end{figure}
\newpage
\begin{figure}
\epsfbox{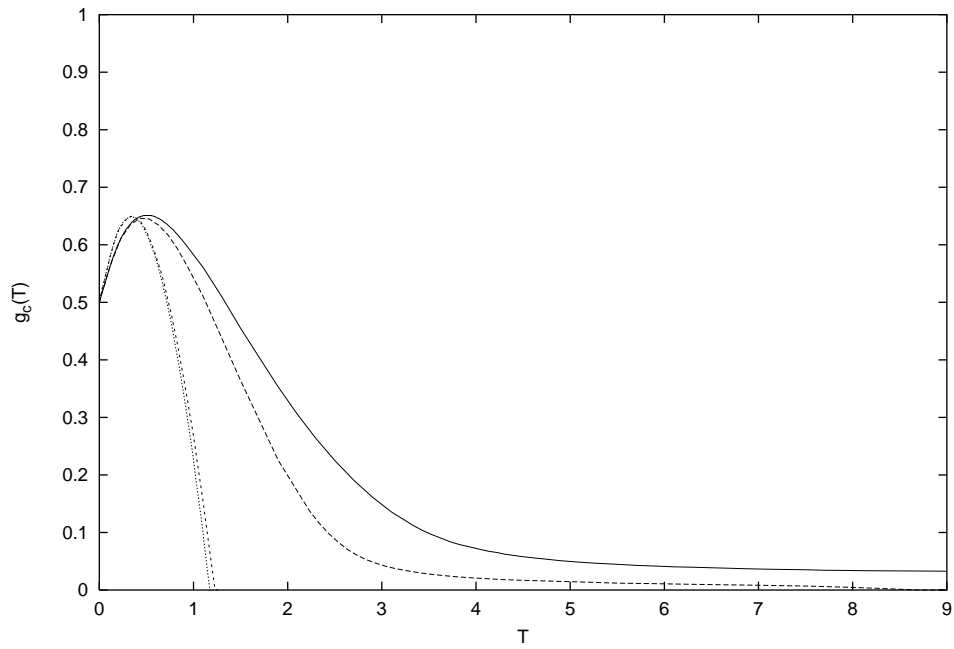}
\\
\caption{Phase diagrams in D=3 and 4 dimensions for models with and without crossing ($\epsilon=1$ in the figure). It can be seen that the difference between the four dimensional models is negligible. In D=3 (model w.o.c.), $T_c\simeq$ 8.49$\epsilon$. 
Solid line: model D=3 w.c.;\\
Long-dashed line: model D=3 w.o.c.;\\
Dashed line: model D=4 w.c.;\\
Dotted line: model D=4 w.o.c.}
\end{figure}
\newpage 
\begin{figure}

\epsfbox{fig3.eps}
\\
\caption {a)We show here an example of Y-shaped configuration for the DNA molecule: in this approximation bubbles are neglected.\\
b)A completely stretched configuration for directed walks, which is dominant for $T\to 0$.}
\end{figure}

\newpage
\begin{figure}

\epsfbox{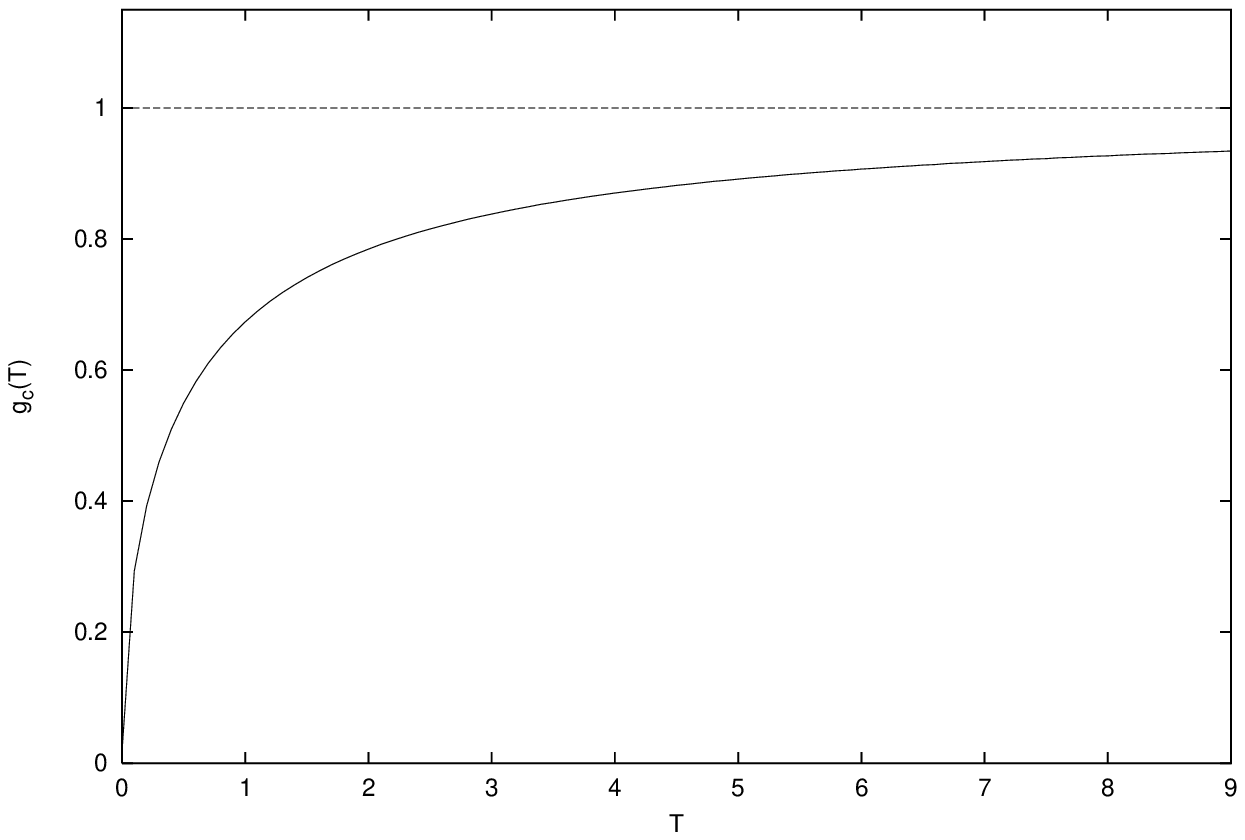}
\\
\caption{Critical force in $d=1$, found with the quantum mapping. The solid line is the result with a symmetric square well, the dashed one with a delta function. The parameters are chosen so that the integral of the potentials over space is the same.}
\end{figure}


\begin{thebibliography}{99}
\bibitem{Car} Note, however, that this is also the case for generic DSAWs, at least in the non-interacting case , where the walks are directed along a generic vector and the self-avoidance constraint is truly effective in the D-1-dimensional transverse space. See {J.~Cardy}, {\it J.~Phys.~A}, {\bf 16}, {L355}, {(1983}).
\bibitem{rw_note} For simple RWs with a force pulling in the direction
$(1,0,\ldots,0)$, the new singularity $z_3$ coming 
from the external force can be obtained 
in the same way as done in Section II. Its value for $d$-dimensional RWs is 
$\frac{1}{2\l(d-1+\cosh{(\beta g)}\r)}$. The value of $z_2$
can be found as in \cite{GB}.
\bibitem{note_1} The exponent 1/2 is given by mean field theory: see \cite{Seb}.
\end{thebibliography}
\end{document}